\documentstyle[floats,aps,psfig]{revtex}

\begin{document}
\draft

\twocolumn[\hsize\textwidth\columnwidth\hsize\csname @twocolumnfalse\endcsname
\title{The Effect of Neutral Atoms on Capillary Discharge Z-Pinch}
\author{R. A. Nemirovsky,\cite{MyEmail} A. Ben-Kish, M. Shuker, and A. Ron}
\address{Department of Physics, Technion, Haifa 32000, Israel.}
\date{\today }
\maketitle
\begin{abstract}
We study the effect of neutral atoms on the dynamics of a capillary
discharge Z-pinch, in a regime for which a large soft-x-ray amplification
has been demonstrated. We extended the commonly used one-fluid
magneto-hydrodynamics (MHD) model by separating out the neutral atoms as
a second fluid. Numerical calculations using this extended model yield
new predictions for the dynamics of the pinch collapse, and better
agreement with known measured data.
\end{abstract}
\vskip2pc] \narrowtext
{\em Z-pinch collapse} has been extensively studied since the late 50s,
being a simple and effective way of producing hot and dense plasma.
In this process, an electric current flowing through a plasma column,
interacts with its self magnetic field,
and the resulting force contracts the plasma column in the
radial direction. Today Z-pinch plasma is widely used
for various applications such as high power radiation sources and neutron
sources \cite{Matzen,Sethian}. An exciting new application of Z-pinch
plasma was recently demonstrated by Rocca et. al.
\cite{Rocca1,Rocca2,Tomasel}. In this work, large amplification of
soft-x-ray light was obtained in Ne-like Ar and S plasma, created by a fast
($\sim 40$ ns) Z-pinch discharge inside a capillary. Compared with the
alternative approach of laser driven excitation \cite{Matthews,Suckewer},
the capillary discharge has the advantage of allowing for compact
({\em table-top}), efficient and simpler soft-x-ray lasers.

In this paper we study the role of neutral atoms in the dynamics of a
capillary discharge Z-pinch, in the regime for which soft-x-ray
amplification was demonstrated. The commonly used one-fluid
{\em magneto-hydrodynamics} (MHD) model assumes that all
the particles in the plasma are charged, and drift together.
We, however, show that for the case discussed here, large portions
of the plasma contain an appreciable amount of neutral atoms.
Since these are not affected by the electro-magnetic forces,
but only by the much weaker mechanical
forces, they flow with much smaller velocities than the ions and the
electrons. To account for this effect, we extend the one-fluid MHD model
by introducing a separate fluid for the neutral atoms
(in addition to the standard electrons-ions fluid). Results of calculations
using this extended model give new predictions for the dynamics of the
pinch collapse, with some features in better resemblance with the measured
data. This confirms our previously reported estimates \cite{Roni}.

We start with the standard one-fluid two-temperature MHD
model, commonly used for numerical calculations of Z-pinch processes
\cite{Shlyaptsev0,Shlyaptsev,Bobrova,Lee-Kim,Lee-Kim1}.
It considers hydrodynamic flow including shock waves,
heat conduction, heat exchange (between ions and electrons), magnetic
field dynamics, magnetic forces, Ohmic heating, radiative cooling and
ionization. We use a simple ionization model,
and assume a quasi steady state, taking into account collisional ionization,
and 2-Body and 3-Body recombination. Since the plasma is assumed to be
optically thin, ionization and excitation by radiation are neglected.
The latter assumption should hold at least to the end of the collapse.
This model is incorporated into our numerical code, SIMBA, where the
equations of motion of the system
(see \cite{Shlyaptsev0,Shlyaptsev,Bobrova,Lee-Kim,Lee-Kim1,Krall})
are solved in a Lagrangean mesh \cite{Morton},
assuming one-dimensional axial symmetry.

Shown to be remarkably stable \cite{Bender},
and having a high length-to-diameter ratio (of 50-500),
the capillary discharge Z-pinch experiment is naturally described in the
framework of this 1-D MHD model. Previously reported
works \cite{Shlyaptsev}, have indicated that taking into account ablation of
the plastic capillary wall is necessary for the correct description of the
pinch dynamics. According to this the calculation should thus be
extended to include a narrow region of the plastic capillary wall. However,
it was also shown in \cite{Shlyaptsev} that even with this effect taken
into account, good agreement with the measured data still requires some
major {\em artificial adjustments} of the plasma transport coefficients.
We have repeated these calculations using the same one-fluid MHD model,
and found them to agree with the reported results.
In particular, we also find that the measured data is reproduced by
one-fluid MHD calculations only when artificial adjustments are introduced,
as demonstrated in Fig. (\ref{fig:1fmodel}).
\begin{figure}[tbh]
\centerline{\psfig{figure=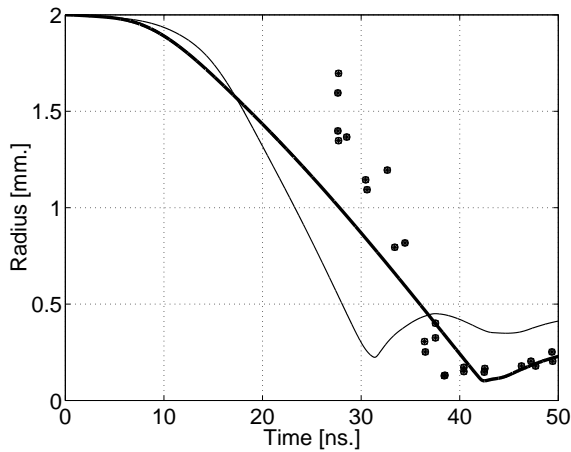,width=3in}}
\vspace{0.3in}
\caption{
\label{fig:1fmodel}
Z-Pinch of Ar plasma inside a plastic capillary. Thin line:
Calculated outer boundary of Ar plasma, assuming classical electrical
conductivity. Thick line: Calculated outer boundary of Ar plasma, with a factor
of 15 on the classical electrical conductivity. Dots: Measured radius of
soft-x-ray source \protect\cite{Shlyaptsev}.
}
\end{figure}
The figure displays the calculated radius of the collapsing Ar plasma as a
function of time in a capillary discharge Z-pinch. The parameters of the
calculations are those used for soft-x-ray amplification experiments
\cite{Rocca1,Rocca2,Shlyaptsev}: initial Ar density of
$\rho _{0}=1.7\cdot 10^{-6}\mbox{g.}/\mbox{cm}^{3}$ or
$\ n_{0}\approx 2.5\cdot 10^{16} \mbox{atoms}/\mbox{cm}^{3}$,
initial temperature of
$T_{0}\approx 0.5$eV, and a maximum current of 39kA, with its peak at
t=32ns \cite{Footnote1}. The figure also presents some measured data,
of the radius of soft-x-ray source, as a function of time,
taken from \cite{Shlyaptsev}. Since the radii of the soft-x-ray source
and that of the collapsing Ar plasma are related,
it is clear that there are disagreements between the calculated
and measured data: For example, the calculated pinch peak is about 10ns
earlier than the measured one. It is shown in Fig. (\ref{fig:1fmodel})
that multiplying the classical electrical conductivity \cite{Braginskii}
by a factor of 15, results in a good agreement with the measured instant
of the pinch peak, however at the same time it also spoils the agreement
with measured collapse velocity. We notice that both calculations do not
properly reproduce the initial stages of the collapse, which is delayed
by about 10-15ns. According to \cite{Shlyaptsev}, reproducing the whole
stages of the measured collapse requires more artificial adjustments in the
plasma transport parameters, up to 20-40 times their classical values.
This need for artificial adjustments of plasma parameters in one-dimensional
one-fluid MHD calculations cannot be explained by two- or three- dimensional
effects in the modeled experiment:
The work of Bender et. al. \cite{Bender} has proven a perfect azimuthal
($\phi$-direction) symmetry in this {\em same} capillary discharge Z-pinch,
and the demonstrated amplification gain \cite{Rocca1,Rocca2,Tomasel}
indicates a very good Z-direction symmetry.

In order to better understand the dynamics of the pinch collapse,
we have focused our study on the importance and the role of neutral atoms
in this process. The one-fluid MHD model assumes that the plasma consists
of two components: electrons and {\em effective} single-type ions, with their
charge being the average charge of all the differently charged ions in
the plasma, including the neutral atoms. In addition, these two components
are assumed to flow together, as a single fluid. This assumptions are
reasonable for regimes for which at least one of the two following
conditions is fulfilled:
({\it i}) All the atoms in the plasma are ionized, or,
({\it ii}) The neutral atoms are strongly coupled to the charged particles,
and hence follow them in the same single fluid.

\begin{figure}[tbh]
\centerline{\psfig{figure=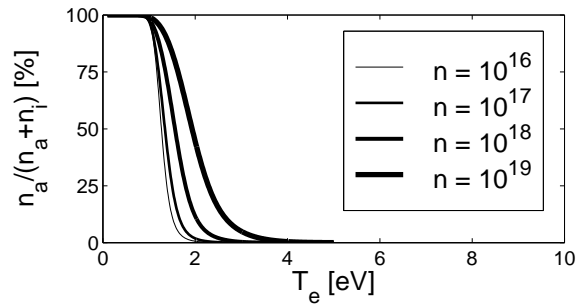,width=3in}}
\vspace{0.2in}
\caption{
\label{fig:arna}
Percentage of neutral atoms in Ar plasma, as a function of electron
temperature. Each line corresponds to a different plasma density.
}
\end{figure}

Fig. (\ref{fig:arna}) presents the percentage of neutral
atoms as a function of electron temperature in Argon plasma,
based on our ionization model.
According to this figure, a plasma of electron temperature lower than 2-3 eV
contains an appreciable amount of neutrals. In Carbon plasma, which is a
typical representative of the ablated capillary wall, the picture is similar.
Our MHD calculations show that the Ar plasma starts to heat up above 2-3 eV
only 5 ns after the beginning of the pinch, and its central region stays
below this temperature for the next 25 ns \cite{Roni}.
Major portions of the plastic wall plasma remain below 2-3eV even after the
pinch collapses at the axis. The percentage of neutral atoms in the
plasma is hence far from being negligible. We thus conclude that condition
({\it i}) does not hold. The plasma contains three different components:
electrons, ions, and neutral atoms. We now turn to check whether or not
condition ({\it ii}) is satisfied, by examining the couplings between these
different ingredients of the plasma.
The electrons and ions, being charged particles, are coupled through
Coulomb forces. A measure of the strength of this coupling is given
by the plasma frequency, $\omega _{P}$. For the case discussed here,
$1/\omega _{P} \approx 10^{-5}-10^{-3}$ns, which is negligible compared to
the typical pinch collapse times of $\tau _{\mbox{pinch}} \approx 40$ns.
This means that the coupling between the electrons and ions is very strong,
and that they practically drift together, as a single fluid.
The neutral atoms, however, are coupled to the charged particles only by
collisions, and may thus flow separately, as a second fluid.
We therefore assume two fluids, one of {\it charged-species} (electrons and ions)
and the other of {\it neutral-species} (atoms),
with flow velocities $u_{i}$ and $u_{a}$ respectively.
The collisional momentum transfer between these two fluids is evaluated
assuming a hard spheres approximation: We regard the two fluids as two
clouds of hard spheres, drifting through one another.
In that case, the collision frequency per unit volume equals:

\begin{equation}
\nu_{ai}^{coll}=\alpha \cdot r_{a}^{2}\cdot n_{a}\cdot
n_{i}\cdot \left| u_{a}-u_{i}\right|  \label{colfreq}
\end{equation}

\noindent and the collisional momentum transfer rate,
per unit volume, is thus

\begin{equation}
F_{ai}^{coll}=\alpha \cdot r_{a}^{2}\cdot m_{a}\cdot n_{a}\cdot n_{i}\cdot
\left| u_{a}-u_{i}\right| \left( u_{a}-u_{i}\right) , \label{colforce}
\end{equation}

\noindent where $\alpha $ is a coefficient of about $2 \pi $.
Here $r$ stands for the particle radius, $m$ for its mass,
and $n$ stands for the number density. The indices $a , i$ denote atoms
and ions respectively. Later on we will use the index $e$ for electrons.

The force in Eq. (\ref{colforce}), $F_{ai}^{coll}$,
depends quadratically on the velocity difference between the
charged-species and the neutral-species fluids.
This coupling thus restrains the separation between the two fluids.
Taking reasonable densities of $n_{a}\approx 10^{16}, n_{i}\approx 10^{15}$
($10\%$ ionization), and an appreciable velocity difference of 
$\left|u_{a}-u_{i}\right| \approx 10^{6}$cm/s, we get for Ar plasma
a collisional coupling term of the order of $10^{6}\mbox{dyn}/\mbox{cm}^{3}$.
This is 2-3 orders-of-magnitude less then the estimated magnetic
($\overrightarrow{j}\times \overrightarrow{B}/c$) and
hydrodynamic ($\nabla P$) forces.

We conclude that in the regime discussed here, both of the above
conditions for the validity of the one-fluid MHD fail to be satisfied.
The two fluids are indeed expected to flow separately. However, they
exchange mass, momentum and energy due to exchange of particles
(by ionization and recombination) and due to atoms-ions collisions.
By $S_a(r,t)$ we denote the mass sink (per unit volume, per unit time)
in the neutral-species fluid due to {\em ionization} of neutral atoms
($S_a \geq 0$). $S_a$ plays a role of a source in the charged-species
fluid. Similarly, $S_i(r,t)$ denotes the mass sink in the charged-species
fluid, due to {\em recombination} of ions$^{+1}$ ($S_i \geq 0$). The total
mass transfer {\em from} the neutral-species fluid {\em into} the charged-species
fluid due to ionization and recombination is thus $S_a-S_i$.

To account for the exchange of mass, momentum and energy between the two
fluids the standard one-fluid MHD for the charged-species fluid
(see \cite{Krall} for example) are amended, and new, separate equations
for the neutral-species fluid are added.
The revised mass equation for the charged-species fluid is then (we use
cylindrical coordinates and assume $\frac{\partial}{\partial\phi}=0 , 
\frac{\partial}{\partial z}=0$):

\begin{equation}
\frac{d}{dt}\left( \rho_{i}+\rho_{e}\right)+\frac{\left( \rho_{i}+%
\rho_{e}\right)}{r}\frac{\partial}{\partial r}\left(ru_{i}\right)=%
\left( S_{a}-S_{i}\right) ,
\label{cmass}
\end{equation}

\noindent where $\rho$ stands for mass density, and
$\frac{d}{dt}\equiv \frac{\partial}{\partial t}+u\cdot\nabla$ is the comoving
derivative. The separate mass equation for the neutral-species fluid is then:

\begin{equation}
\frac{d}{dt}\left( \rho_{a}\right)+\frac{\rho_{a}}{r}%
\frac{\partial}{\partial r}\left(ru_{i}\right)=%
-\left( S_{a}-S_{i}\right)
\label{nmass}
\end{equation}

\noindent The revised momentum equation for the charged species fluid is:
\begin{eqnarray}
\left( \rho _{i}+\rho_{e}\right)\frac{d}{dt}u_{i}=%
-\frac{\partial}{\partial r} \left( P_{e}+P_{i}\right)+%
\frac{\overrightarrow{j}\times \overrightarrow{B}}{c} \nonumber \\
+ F_{ai}^{coll} + S_{a}\left( u_{a}-u_{i}\right) ,
\label{cmome}
\end{eqnarray}

\noindent where $P$ stands for pressure, $\overrightarrow{j}$ for current
density and $\overrightarrow{B}$ for magnetic field. $F_{ai}^{coll}$ is
the collisional momentum exchange between the neutral-species fluid and the
charged-species fluid, given in Eq. (\ref{colforce}).
\noindent The momentum equation of the neutral-species fluid should then be:

\begin{equation}
\rho _{a}\frac{d}{dt}u_{a}=-\frac{\partial}{\partial r} \left( P_{a}\right)%
- F_{ai}^{coll} + S_{i}\left( u_{i}-u_{a}\right)
\label{nmome}
\end{equation}

\noindent Similarly, the one-fluid MHD ion-energy equation
\cite{Shlyaptsev0,Shlyaptsev,Bobrova,Lee-Kim,Lee-Kim1})
is also properly amended, and separate atom-energy equation for
the neutral-species fluid is introduced. Collisions between the two
fluids, as well as particles exchange due to ionization and
recombination are considered in these equations in the same manner as
in the mass and momentum equations. The MHD electron-energy equation is
left unchanged.

\begin{figure}[tbh]
\centerline{\psfig{figure=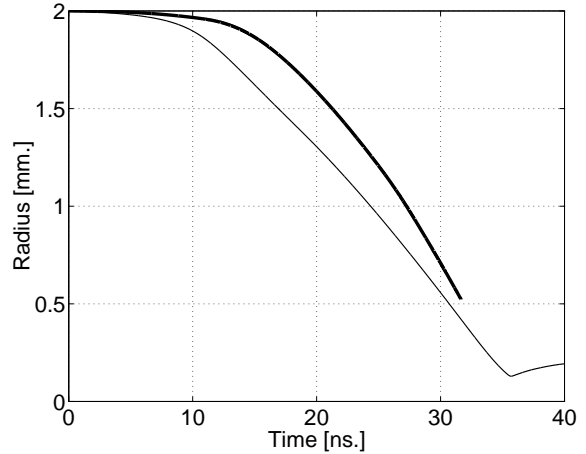,width=3in}}
\vspace{0.3in}
\caption{
\label{fig:1f-2frhis}
Calculated outer boundary of Ar plasma during a Z-pinch capillary discharge.
Thin line: Neutral atoms neglected (standard one-fluid MHD).
Thick line: Neutral atoms included (extended model).
}
\end{figure}

These equations were incorporated into our SIMBA code.
For simplicity, and in order to emphasize the effect introduced by
separating the neutral atoms from the charged-species fluid,  we assume, in
the following calculations, that the capillary wall is also made of Argon.
The other pinch parameters are left unchanged, however we now use
the classical transport coefficients \cite{Braginskii},
without any artificial adjustments.
Fig. (\ref{fig:1f-2frhis}) shows the effect of the neutral-species fluid on
the calculated outer boundary of the collapsing Ar plasma.
It is clearly indicated that the effect of the neutral component in the
capillary discharge Z-pinch is not negligible. When the
neutral-species fluid is included the collapse seems to be delayed, however
after it starts it is more rapid. This trend seems to better resemble the
data presented in Fig. (\ref{fig:1fmodel}), where it was shown that compared
to one-fluid MHD calculations the measured collapse is delayed,
and after it starts the collapse rate is much higher.

\begin{figure}[tbh]
\centerline{\psfig{figure=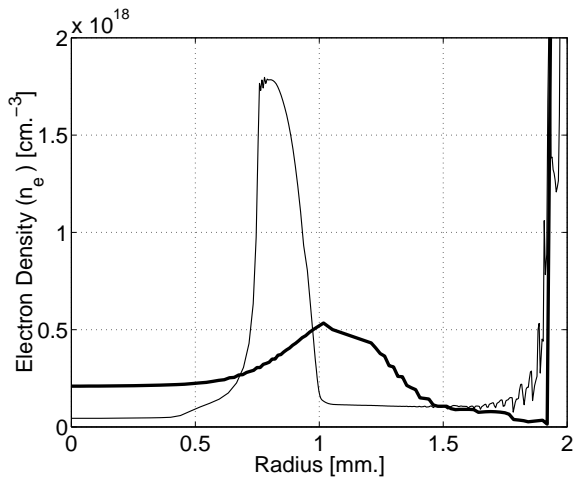,width=3in}}
\vspace{0.3in}
\caption{
\label{fig:1f-2fprof}
Calculated electron density profiles at time=25ns. of an Ar Z-pinch capillary
discharge. Thin line:
Neutral atoms neglected (standard one-fluid MHD).
Thick line: Neutral atoms included (extended model).
}
\end{figure}

We have also examined the effect of neutral atoms on the electron density
distribution during the pinch. In Fig. (\ref{fig:1f-2fprof}), the
calculated spatial distribution of electron density at time=25ns is plotted,
with and without the neutral-species fluid. Both models predict a collapsing
plasma sheath, and show some ablated material from the capillary wall.
However, when the neutral-species fluid is taken into account,
the collapsing plasma sheath is wider and less dense,
compared to the predictions of the standard one-fluid MHD model.

We like to offer a qualitative explanation for the results presented in
Fig. (\ref{fig:1f-2frhis}),(\ref{fig:1f-2fprof}).
In the one-fluid MHD model, the atoms and ions are assumed to flow together
with the electrons. The magnetic forces, which
are dominant in this case, thus accelerate the whole plasma body.
In reality, however, only the ions flow together with the electrons, while
the neutral atoms flow separately. Since the plasma is initially mostly
neutral, the magnetic forces act only on a small fraction of the total
mass, which is then rapidly accelerated inwards.
Most of the Ar stays outside, almost at rest. While the process evolves,
more atoms get ionized, and join the charged-species fluid. This effect is
seen in Fig. (\ref{fig:1f-2frhis}) as a delay in the collapse.
At any given spatial and temporal point, the magnetic forces act on a 
``freshly'' ionized matter, almost at rest. The resulting acceleration
is thus more gradual, leading to a wider and less dense plasma sheath,
as seen from Fig. (\ref{fig:1f-2fprof}).

In conclusion, we have shown that the effect of neutral atoms on the
dynamics of the capillary discharge Z-pinch is not negligible.
We have demonstrated that separating out the neutral atoms as a second fluid
produces a different pinch collapse dynamics, with some features similar to
the measured data. It is expected that the improved modeling of the pinch
collapse dynamics will yield a better understanding of capillary discharge
X-ray lasers, since the amplification gain, as well as the propagation and
refraction of radiation in the lasing media are both dominated by the
details of the plasma state.

{\em Acknowledgments}: We gratefully acknowledge the help of A. Birenboim,
J. Nemirovsky, and J. Falcovitz for their advice and useful suggestions.
This work was partially supported by the Fund for Encouragement of
Research in the Technion.


\begin{references}
\bibitem{Footnote1}  A detailed description of initial and boundary
conditions in one-fluid MHD calculations of this same capillary discharge
Z-pinch is given in \cite{Bobrova}.
\end{references}
\end{document}